\documentclass[fleqn,10pt]{wlscirep}
\usepackage[utf8]{inputenc}
\usepackage[T1]{fontenc}
\usepackage{subcaption}
\usepackage{float} 
\usepackage{color}
\usepackage{array}            
\usepackage{threeparttable}  
\usepackage{adjustbox}       
\usepackage[table]{xcolor}
\usepackage[section]{placeins}

\title{Hidden multistate models to study multimorbidity trajectories}
\author[1]{Valentina Manzoni}
\author[1,2]{Francesca Ieva}
\author[3,4]{Amaia Calderón-Larra\~naga}
\author[3,4]{Davide Liborio Vetrano}
\author[3*]{Caterina Gregorio}
\affil[1]{MOX - modeling and Scientific Computing Laboratory, Department of Mathematics, Politecnico di Milano, Italy}
\affil[2]{HDS, Health Data Science center, Human Technopole, Italy}
\affil[3]{Aging Research Center, Department of Neurobiology, Care Sciences and Society, Karolinska Institutet and Stockholm University, Sweden}
\affil[4]{Stockholm Gerontology Research Center, Stockholm, Sweden}

\affil[*]{caterina.gregorio@ki.se}

\begin{abstract}

Multimorbidity in older adults is common, heterogeneous, and highly dynamic, and it is strongly associated with disability and increased healthcare utilization. However, existing approaches to studying multimorbidity trajectories are largely descriptive or rely on discrete-time models, which struggle to handle irregular observation intervals and right-censoring. We developed a continuous-time hidden multistate modeling framework to capture transitions among latent multimorbidity patterns while accounting for interval censoring and misclassification. A simulation study compared alternative model specifications under varying sample sizes and follow-up schemes, and the best-performing specification was applied to longitudinal data from the Swedish National study on Aging and Care–Kungsholmen (SNAC-K), including 2,716 multimorbid participants followed for up to 18 years. Simulation results showed that hidden multistate models substantially reduced bias in transition hazard estimates compared to non-hidden models, with fully time-inhomogeneous models outperforming piecewise approximations. Application to SNAC-K confirmed the feasibility and practical utility of this framework, enabling identification of risk factors for accelerated progression toward complex multimorbidity and revealing a gradient of mortality risk across patterns. Continuous-time hidden multistate models provide a robust alternative to traditional approaches, supporting individualized predictions and informing targeted interventions and secondary prevention strategies for multimorbidity in aging populations.

\end{abstract}

\begin{document}

\flushbottom
\maketitle
%
%
\thispagestyle{empty}

\section*{Introduction}

The global population is aging at an unprecedented rate, with projections indicating that by 2050, individuals aged 60 and over will outnumber younger age groups in many regions of the world \cite{UN_pop_prospect}. This demographic shift brings profound implications for public health, as aging is increasingly associated with complex, dynamic, and heterogeneous health trajectories. Such patterns challenge traditional approaches that focus on single diseases, linear aging processes, or short-term clinical outcomes, and highlight the need for models capable of capturing interdependent and evolving health changes.
To address this complexity, there is a growing need for advanced statistical frameworks that can represent the intrinsic heterogeneity of aging processes over time. These tools are essential for studying geriatric syndromes, which inherently reflect the multifactorial and systemic nature of aging. Among them, multimorbidity—defined as the co-occurrence of multiple chronic conditions \cite{Busija2019}—has emerged as the most prevalent and impactful health issue in older adults \cite{Barnett2012Multimorbidity}. Multimorbidity serves as a paradigmatic example of the aging process itself, capturing the cumulative, interacting, and individualized patterns of health decline that characterize later life.
Multimorbidity has significant clinical relevance.  Among older adults, 10–15\% experience physical frailty \cite{Tazzeo2021}, 20\% will receive a diagnosis of dementia \cite{Grande2021}, and 15–20\% will become dependent in activities of daily living \cite{Marengoni2021}. Multimorbidity significantly drives healthcare utilization, accounting for the majority of general practitioner consultations and increasing the risk of hospital admissions \cite{Romana2020}. 
A defining feature of multimorbidity is that multiple diseases do not occur independently, but instead co-occur in patterns that exceed random chance \cite{prados}. These clusters can arise from shared environmental exposures, causal relationships between conditions, adverse effects of treatments, or common genetically determined mechanisms \cite{Langenberg2023}. 
Since simple disease counts fail to reflect the complexity of multimorbidity, pattern-based approaches are increasingly used to identify more homogeneous subgroups. These patterns have shown predictive value for outcomes such as mortality \cite{Vetrano2021,CalderonLarranaga2017} frailty \cite{Tazzeo2021}, disability, institutionalisation, and hospitalizations \cite{Vetrano2021}, supporting their potential use in guiding targeted prevention and care strategies.
While pattern-based approaches have advanced our ability to stratify older populations and predict adverse outcomes, they remain largely static and cross-sectional. A major challenge in the field is how to move beyond these snapshots to characterize the trajectories of multimorbidity over time, thereby capturing its dynamic and evolving nature \cite{Cezard2021,Nagel2024}. Understanding these trajectories is crucial for (1) identifying risk factors that drive progression toward more clinically complex multimorbidity patterns, and (2) recognizing homogeneous longitudinal paths with shared prognoses and care needs \cite{CalderonLarranaga2025}. However, the main barrier to generating such evidence is the absence of tailored statistical modeling frameworks capable of addressing these longitudinal and multidimensional research questions. In the realm of longitudinal methods for studying multimorbidity, multistate modeling offers a promising approach to capture how disease patterns evolve over time \cite{Nagel2024}. These models extend survival analysis by allowing transitions across a set of intermediate states and an absorbing state, typically death. However, traditional multistate models often assume that these states are directly observable, which is rarely the case in multimorbidity research, where disease patterns are latent and prone to misclassification.
To address this limitation, we propose a flexible framework based on continuous-time Hidden Markov Models (HMMs). This approach retains the strengths of multistate models—such as handling irregular observation times, incorporating covariates, and modeling time-varying transition risks—while explicitly accounting for uncertainty in state classification. It also remains parsimonious in terms of parameter estimation, making it suitable for complex longitudinal data.
We illustrate the utility of this framework through a simulation study that explores how different model specifications perform under varying study designs and data conditions. Finally, we apply the methodology to real-world data from the Swedish National study on Aging and Care–Kungsholmen (SNAC-K), examining how socioeconomic and lifestyle factors influence transitions to more severe multimorbidity states. This application demonstrates the practical relevance of continuous-time HMMs as a robust tool for modeling multimorbidity trajectories in aging research.

\section*{Methodology}

\begin{figure}[h]
    \centering
    \includegraphics[width=\textwidth]{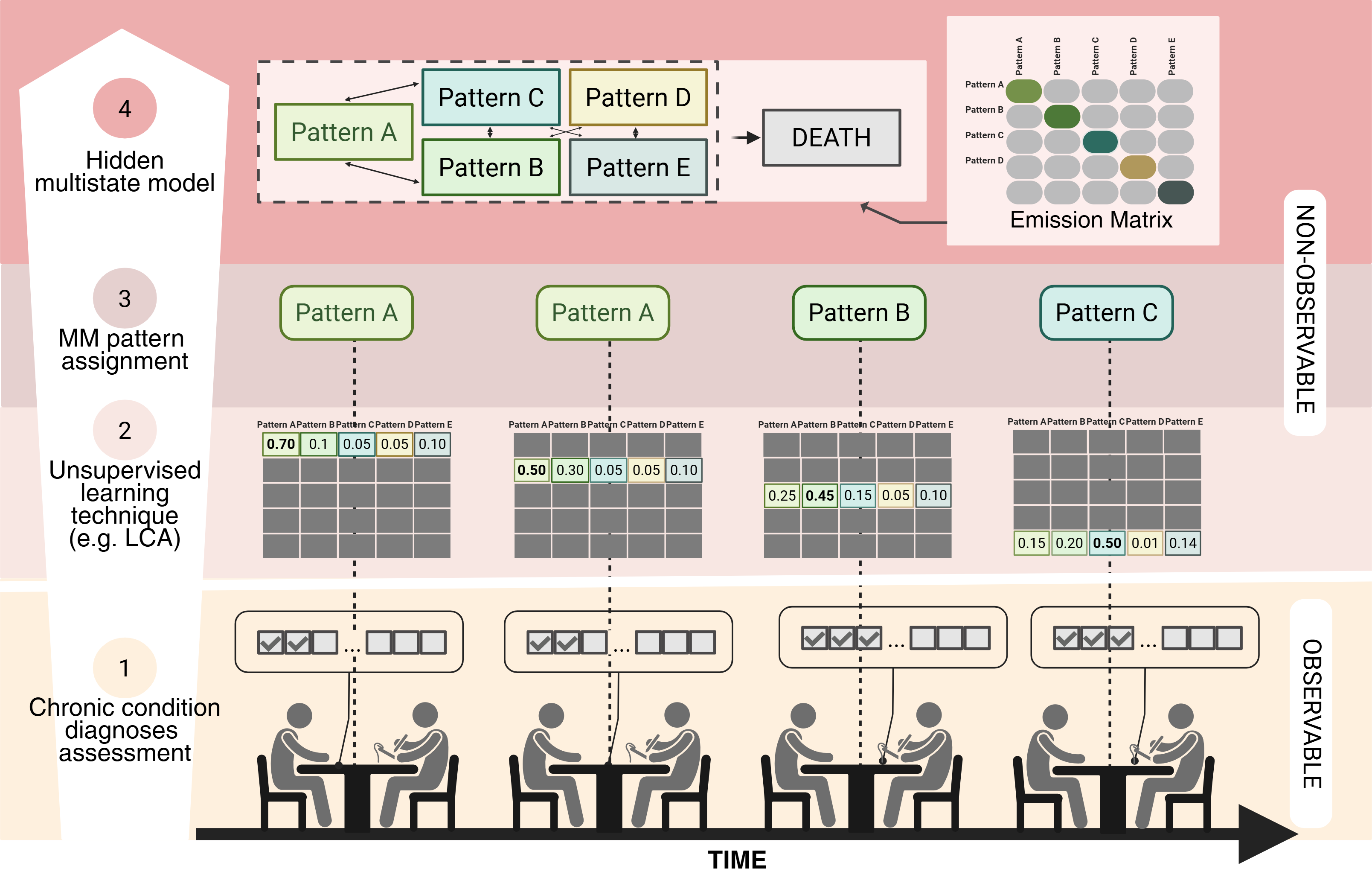}
    \caption{\textbf{ Conceptual and analytical pipeline for modeling multimorbidity trajectories.} Chronic disease diagnoses are first aggregated using Latent Class Analysis to identify multimorbidity patterns (Steps 1–2). Individuals are assigned pattern membership at each visit (Step 3), and transitions across latent states—including progression to death—are modeled using a continuous‑time hidden multistate model that accounts for misclassification and interval censoring (Step 4). \small{LCA: Latent Class Analysis; MM: Multimorbidity. Created in BioRender. } }
    \label{fig:main}
\end{figure}

We begin by introducing the multistate modeling framework in the context of chronic disease research, which serves to highlight the key challenges and distinctive features of applying such models in this domain. We then extend the framework to accommodate transitions across latent multimorbidity patterns, which represent clusters of co-occurring diseases. Figure \ref{fig:main} presents an overview of the proposed methodology and conceptual framework, offering a graphical summary of our approach.

\subsection*{Multistate Modeling Framework}

A multistate model describes a stochastic process, $S(t), \, t > 0$, that records subjects' moving across a set of discrete health states, $\mathcal{S}$ \cite{Andersen2002Multi-stateAnalysis.,Hougaard2000DependenceStructures}, over time. $\mathcal{S}$ contains a number of transient states, reflecting different health-living states, and "Death" as the absorbing state, i.e. subjects cannot exit this state and transition to others. Time is treated as a continuous variable, allowing transitions between states to occur at any moment—a realistic assumption when modeling medical and biological processes. Specifically, the timescale used in the multistate models considered here is the individual's age. Consequently, each subject may have a different time origin in the model, corresponding to their age at study entry.

In the context of aging and the development of chronic conditions, it is highly unlikely that transitions between health states are recorded at the exact moment they occur. Only acute health events typically allow for precise timing of state changes. More commonly, individuals are assessed at discrete intervals—such as during routine medical visits or scheduled health surveys—and transitions are inferred from these periodic observations. This observation mechanism results in panel data, characterized by intermittent observation and interval censoring: while it is known that a transition occurred between two observation points, the exact timing remains unknown. In addition, unlike continuous monitoring, which captures every state transition in real time, panel data provide only snapshots of an individual's health status at specific time points. As a result, the complete sequence of states occupied during the observation period is often partially observed, and some transitions may go undetected. Traditional multistate models, however, assume continuous observation and precise tracking of state changes, which may not align with the realities of such data.

To address this, our work employs multistate modeling approaches specifically designed for panel data, which account for the intermittent nature of observations during parameter estimation. Importantly, we assume that observation times are non-informative—that is, the timing of measurements is independent of the underlying multistate process. Additionally, we adopt the Markov assumption, meaning that the probability of transitioning to a future state, and the timing of that transition depend solely on the current state and the individual's current age:

\begin{equation}
\Pr[S(t+h) = y \mid S(r) = s(r), \, r \leq t] = \Pr[S(t+h) = y \mid S(t) = S(t)], \quad \forall h > 0.
\end{equation}

A multistate model is uniquely defined by a set of transition intensity functions, \(q_{ij}(t, x)\),  which represent the instantaneous probability at each time $t$ of moving between each pair of states \(i\) and \(j\), for \(i\), \(j\) $\in \mathcal{S}$.  In the continuous-time case, the transition intensities can be represented in a \(nxn\) matrix \(Q\), with \(n\) total number of states, whose rows sum to zero.
The Kolmogorov Forward Equations (Chapman-Kolmogorov Differential Equations) are used to describe the evolution of probabilities over time in a Continuous Time Multistate Models (CTMM). These equations provide a mathematical framework to calculate the probability of being in a specific state at any given time. Let \( P(t) \) represent the \( n \times n \) matrix of state probabilities, where \( p_{ij}(t) \) is the probability of being in state \( j\) at time \( t \), given that the process started in state \( i \) at time 0. The Kolmogorov Forward Equation is given by:

\begin{equation}
\frac{dP(t)}{dt} = P(t) \cdot Q
\end{equation}

For each state pair \( (i, j) \), the derivative \( \frac{dp_{ij}(t)}{dt} \) represents the instantaneous rate of change of the probability of being in state \( j \) at time \( t \), starting from state \( i \). In the context of aging and chronic diseases, the matrix, \(Q(t)\), of the transition intensities needs to be allowed to vary with time, allowing the probability of transitions across the state to vary as the individual ages. As a result, the solution of the Kolmogorov Forward Equations (KFEs) becomes more complex as it requires integration over time. Direct integration requires the use of numerical solvers \cite{LSODA}, which can be more computationally intensive and time-consuming. Alternatively, a numerical approximation that uses a piecewise constant version of the \( Q(t) \) matrix can be used to simplify computations. To approximate the solution in the time-inhomogeneous case, the time interval \([0, t]\) can be divided into small subintervals \([t_i, t_{i+1}]\), where \( \Delta t = t_{i+1} - t_i \) is the step size. The transition intensity matrix \( Q(t) \) is approximated as piecewise constant within each interval. This leads to the following approximation of the solution:

\begin{equation}
P(t) \approx P(t_0) \cdot \prod_{i=0}^{n-1} e^{\Delta t \cdot Q(t_i)}
\end{equation}

A crucial step in the modeling process involves selecting a parametric form for the intensity function. We adopt the proportional hazards model due to its interpretability and widespread use in clinical research. While alternative parametric families could theoretically be considered, this study focuses on the Gompertz distribution, which reflects the natural tendency of transition intensities—such as progression to greater disease burden or death—to change monotonically with age. This choice aligns with well-documented patterns in health and disease progression \cite{Riggs1990, Zamsheva2024, Kuss2024}.
According to this parametric proportional-hazard model, covariates $\textbf{x}$ associated with transition intensities are assumed to have a multiplicative effect, and each transition $i \to j $ can have a different set of covariate effects. The transition intensities take the form:
\begin{equation}
    h_{ij}(t | x) = \lambda_{ij} e^{\alpha_{ij} t + \beta_{ij}^T \textbf{x}}
\end{equation}

where \(\lambda_{ij}\) is the baseline rate parameter in exponential form for the transition from state \(i\) to \(j\), \(\alpha_{ij}\)
the shape parameter for the transition from state \(i\) to \(j\), \(\textbf{x}\) the vector of covariates and \( \mathbf{\beta_{ij}}\) the vector of regression coefficients.
When \(\alpha<0\), the hazard decreases over time, whereas \( \alpha>0\) characterizes an increasing hazard with age. For \( \alpha=0\), the Gompertz model reduces to the exponential model, i.e. a time-homogeneous model.

\subsection*{Hidden Multistate Models for Latent Multimorbidity Patterns}

Multimorbidity states are conceptualized as latent constructs—unobservable directly but inferred from a set of categorical disease indicators, denoted as
$\textbf{Y}(t) = (Y_1(t), Y_2(t), \dots, Y_R(t))$,
where each $Y_r(t)$ indicates the presence or absence of a specific chronic condition at age $t$. The probability that an individual  $k$  belongs to a particular multimorbidity pattern, conditional on the diseases developed by time $t$, is given by
$P(C_k(t) = c \mid \textbf{Y})$,
where $c \in \{1, 2, \dots, C\}$. This probability can be estimated using existing unsupervised learning techniques, such as Latent Class Analysis (LCA) or other soft-clustering methods.  Let \(\hat{P}(C_k (t) = i \mid \textbf{Y})\) denote the posterior probability that the individual \(k\) belongs to the latent class \(i\) estimated from the latent class model. 

Based on the observed values $\textbf{y}(t)$, an individual can be assigned to the most likely pattern using this probability, resulting in the observed state $W(t)$. However, when analyzing transitions between latent multimorbidity states, it is important to recognize that only $W(t)$ is observed, not the true latent state $C(t)$. This problem can be framed as a hidden multistate model that extends the classical framework by explicitly modeling the generation of the observed states from the latent (hidden) ones \cite{miscMarkov}. For each individual $k$, at observation time $t_{kn}$, the observed states $W$ are generated conditionally on the latent states $C$ via an emission matrix $E$. This is an $n \times n$ matrix, where the entry $(i, j)$ represents the probability of observing state $j$ given that the hidden state is $i$.  The emission probabilities are defined as:

\begin{equation}
    e_{i,j} = P(W(t) = j \mid C(t) = i) 
    \label{eq:emission_prob}
\end{equation}

In practice, these probabilities are estimated using posterior class-membership probabilities obtained from the latent class model, as defined above.  Then, the emission probabilities can be estimated as \cite{Vermunt2010}:
\[
\hat{e}_{i,j}
=
\frac{\sum_{k=1}^N \mathbf{1}(W_k = j)\, \hat{P}(C_k = i \mid Y_k)}
{\sum_{k=1}^N \hat{P}(C_k = i \mid Y_k)}.
\]

where $N$ is the total number of individuals in the sample. 
This estimator approximates the joint distribution \(P(W, C)\) using posterior expectations and normalizes to obtain \(P(W \mid C)\). The emission matrix is assumed to be time-invariant, as it depends only on the measurement model derived from the baseline latent class analysis. \\

Since exact states are unknown, subject $k$'s contribution to the likelihood\cite{msm} needs to be calculated over all possible paths of underlying states $C_{k1}, \ldots, C_{k n_k}$:

\begin{equation}
L_k = \Pr(w_{k1}, \ldots, w_{k n_k}) 
= \Pr(w_{k1}, \ldots, w_{k n_k} \mid C_{k1}, \ldots, C_{k n_k}) \Pr(C_{k1}, \ldots, C_{k n_k})
\end{equation}

Assuming that the observed states are conditionally independent given the values of the underlying states and the Markov property, the contribution $L_k$ can be decomposed into sums over each underlying state. The sum is accumulated over the unknown first state, the unknown second state, and so on until the unknown final state:

\begin{align}
L_k = \sum_{C_{k1}} \ldots \sum_{C_{k n_k}} 
& \Pr(W_{k1} \mid C_{k1}) \Pr(C_{k1}) \cdot \Pr(W_{k2} \mid C_{k2}) \Pr(C_{k2} \mid C_{k1}) \cdots \Pr(W_{k n_k} \mid C_{k n_k}) \Pr(C_{k n_k} \mid C_{k n_k-1}),
\end{align}

where \( \Pr(W_{kn} \mid C_{kn}) \) is the emission probability from the hidden state \( C_{kn} \) to the observed state \( W_{kn} \).  The emission probabilities are treated as fixed quantities, estimated in the preliminary step as explained above, and then incorporated into the likelihood.  The term \( \Pr(C_{k,j+1} \mid C_{kj}) \) is the \( (C_{kj}, C_{k,j+1}) \)-th entry of the Markov chain transition probability matrix \( P(t) \), evaluated at \( t = t_{k,j+1} - t_{kj} \). If the hidden state is death, measured without error, whose entry time is known exactly, then the contribution to the likelihood is summed over the unknown state at the previous instant before death.

\FloatBarrier 
\section*{Simulation Study}

Simulation studies use pseudo-random sampling to generate data. They constitute an invaluable tool for statistical research, particularly for the evaluation of new methods and for the comparison of alternative methods. By simulating data under known conditions, the data-generating mechanism is defined by the researchers, making it possible to observe if the analysed methods are able to recover such a  "ground truth". In the following, we use the ADEMP structure  \cite{ADEMP_framework} 
to present the details of the simulation study considered in this work.

The goal of this simulation study is to 1) verify the need to consider the latent nature of the states when modeling transition among multimorbidity patterns and, 2) compare different specifications of HMMs under different scenarios (e.g. study type, sample size).

\subsection*{Data-generating Mechanism}

\begin{figure}[ht]
    \centering
    \includegraphics[width=1\linewidth]{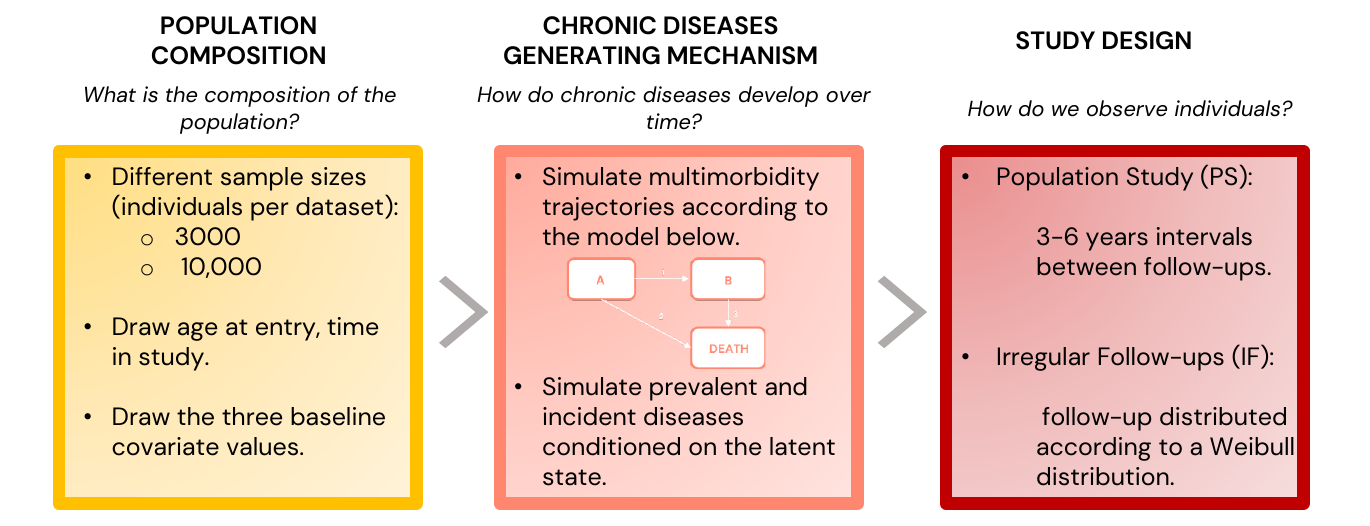}
    \caption{\textbf{Overview of the data‑generating mechanism used in the simulation study.} The process integrates population composition, chronic disease development, and observation schemes, with scenarios varying by sample size (3,000 vs. 10,000) and follow‑up structure (regular population‑based intervals vs. irregular visits). Transitions among multimorbidity states and death follow a Gompertz‑based continuous‑time process.}
    \label{fig:data_gen}
\end{figure}

All key simulation parameters—including the cohort's age range, disease prevalences, number of latent classes, types and number of simulated diseases, minimum and maximum ages at study entry, and intervals between follow-up visits—are informed by the application data. This approach is taken to ensure that the data-generating process closely reflects realistic conditions.
Figure \ref{fig:data_gen} presents the three main components of the data-generating process along with the four simulation scenarios considered. A detailed step-by-step description of the data generation procedure is provided in the supplementary materials. For each scenario, 100 datasets are generated. The scenarios differ in terms of sample size (3,000 vs. 10,000 individuals per dataset) and observation scheme (studies with regular vs. irregular follow-ups ).

It is important to note that for illustrative purposes, the simulation study is conducted in a simplified setting in which chronic diseases are grouped into two different multimorbidity states: multimorbidity pattern A and multimorbidity pattern B. In addition to these states, an absorbing state representing "Death" is included, with the exact time of transition to this state assumed to be known. Consequently, the data-generating multistate model comprises three states: (1) Multimorbidity Pattern A, (2) Multimorbidity Pattern B, and (3) Death. The model allows for three possible transitions: from Pattern A to Pattern B (Transition 1), from Pattern A to Death (Transition 2), and from Pattern B to Death (Transition 3).

Transition intensities between states are modeled using a Gompertz distribution. Three covariates are included in the simulation, although only two are specified to influence the transition intensities. The hazard function for the Gompertz distribution, which governs the transition rates, is defined as follows:

 \begin{equation}
     q_{ij}(t) = \lambda_{ij} \exp(\alpha_{ij} t) \exp( \beta_{ij.1} x_1 + \beta_{ij.2} x_2)
 \end{equation}   

And the loglinear formulation:
\begin{equation}
    \ln q_{ij}(t | x) = \ln(\lambda_{ij}) + \alpha_{ij} t + \beta_{ij.1} x_1 + \beta_{ij.2} x_2
\end{equation}

The parameter values are reported in Table \ref{tab:True parameters' values for simulation} according to the loglinear formulation.

\begin{table}[ht]
\centering
\begin{tabular}{|c|l|l|l|l|l|l}
\hline
\textbf{Transition} & \textbf{Rate $log(\lambda)$} & \textbf{Shape $\alpha$} & \textbf{$\beta^{x_1}$} & \textbf{$\beta^{x_2}$}  & \textbf{$\beta^{x_3}$} \\
\hline
1 & -14.79613 & 0.1556735 & -0.1957039 & -0.2403646  & 0\\
2 & -11.90339 & 0.1088540 & 0.1405208 & -0.4468561 & 0\\
3  & -11.55569 & 0.1071397 & 0.1860589 & -0.3101002 & 0\\
\hline
\end{tabular}
\caption{True multistate model parameters from which the simulated data are generated.}
\label{tab:True parameters' values for simulation}
\end{table}

Prevalent conditions at baseline are simulated conditionally on the individual’s initial multimorbidity pattern. Incident diseases are then simulated from the set of conditions not yet developed by the individual, based on the state toward which they are transitioning. If the next transition is to the absorbing state of death, incident diseases are simulated conditionally on the current state instead. Additionally, we assume the presence of a subset of rare diseases (with a theoretical baseline prevalence of less than 2\%) that occur independently of the multimorbidity patterns. Following disease simulation, study design schemes—such as population-based sampling or irregular follow-ups—are applied to mimic the data collection process in real-world longitudinal studies. As a result, disease onset is only recorded at the first follow-up visit after the actual onset, reflecting the interval-censored nature of observational data.

\subsection*{Analysis of the simulated data}

The following models are presented and compared in the simulation study:
\begin{enumerate}
    \item Non-hidden multistate model with approximate Gompertz baseline (\textbf{ApproxTIMM}). A piece-wise constant baseline hazard function is used to approximate the time-inhomogeneous Gompertz baseline. The baseline hazard function is calculated over a set of time points covering the desired age range. The shape and rate parameters are then determined using linear interpolation. The model accounts for interval-censored data, where the exact timing of a transition is unknown but is known to have occurred within a specific time window between follow-up visits. The model is fitted with the R package \texttt{msm} \cite{msm}.
    
    \item Hidden multistate model with approximate Gompertz baseline (\textbf{ApproxTIHMM}). This model extends the ApproxTIMM by considering that the states are latent. The model is fitted with the \texttt{msm} package.

    \item Non-hidden time-inhomogeneous multistate model and Gompertz baselines (\textbf{TIMM}). TIMM uses direct numerical solutions of the system of differential equations to compute the likelihood, providing a more precise and accurate representation of the transition dynamics. However, the computational demands are significantly higher due to the complexity of solving differential equations numerically. It is implemented with \texttt{nhm} R package \cite{nhm}.
    
    \item Hidden time-inhomogeneous multistate model with Gompertz baseline (\textbf{HTIMM}). This model extends the TIMM by incorporating latent states. Implemented with \texttt{nhm} R package. 
\end{enumerate}

Model performance is evaluated against a benchmark model (REF), which serves as a hypothetical best-case comparator. This model reflects an idealised scenario—unattainable in real-world settings—in which latent states are fully observable and exact transition times are known. The benchmark model employs a Gompertz hazard function, consistent with the specification used in the data-generating process, and is estimated using the \texttt{flexsurv} package \cite{flexsurv}.

The performance of the methods in estimating model parameters is evaluated based on the average of the estimates, the standard deviation of the estimates, the average of the standard errors, the average bias, and the coverage of confidence intervals.


To validate the data-generating mechanism of the simulation, we first evaluated the performances of the benchmark model (REF), which assumes exact knowledge of both transition times and states.
As expected, this model correctly recovers the true parameters, as illustrated in Table \ref{tab: transition 1}, Table \ref{tab: transition 2}, and Table \ref{tab: transition 3},  confirming the validity of the data-generating mechanism. Furthermore, as the sample size of subjects in the datasets increases, so does the accuracy of the estimates for the benchmark model, while their bias decreases.

Table \ref{tab: transition 2} and Table \ref{tab: transition 3} show that all model specifications adequately recovered the covariate effects for Transition 2 and 3 into the absorbing non-latent state (Death, state 3). However, performance measures decreased for transitions between the two states of multimorbidity (Table \ref{tab: transition 1}), where misclassification can occur due to the latent nature of the states. In these cases, the hidden Markov models (ApproxTIHMM and TIHMM) outperformed their non-hidden counterparts (ApproxTIMM and TIMM), showing lower bias and higher coverage.

When comparing different study designs, the irregular visits scenario (median number of follow-ups per subject: 6, interquartile range: 3-10)  exhibits lower bias but also reduced coverage compared to the population-based study scenario (median number of follow-ups per subject: 2, interquartile range: 2-3). This seemingly paradoxical outcome arises from a decrease in standard errors, while small biases persist, ultimately leading to slight undercoverage. A similar pattern can be observed when comparing scenarios with varying sample sizes. For instance, in simulations with a larger number of individuals (Nsim = 10,000), model estimates show reduced coverage due to the same reduction in standard error.

Regarding the estimation of the baseline transition hazard parameters (shape and rate), the fully time-inhomogeneous models (TIMM and TIHMM) demonstrated better performance than those using piecewise constant approximation of the baseline. The latter showed more variability and scenario-dependent behavior (see supplementary material), indicating that such approximations may inadequately capture the true underlying hazard dynamics in time-varying settings. 

Overall, simulation results indicate that the fully time-inhomogeneous model (TIHMM), which integrates interval censoring and the latent nature of the states, is better suited to accurately modeling transitions among latent multimorbidity patterns.
In addition, the empirical standard deviations of the estimates across simulation replicates were generally in close agreement with the corresponding average model-based standard errors, indicating good calibration of the inferential procedure.

\begin{table}[ht]
\caption{Simulation study results for Transition 1 (From Mild MM to Complex MM)}
\begin{centerbox}
\begin{threeparttable}
\setlength{\tabcolsep}{6pt}
\begin{tabular}{l c c c c c | c c c c c}

\hline
 & \multicolumn{10}{c}{Population Study} \\
\hline
 & \multicolumn{5}{c}{n = 3000} & \multicolumn{5}{c}{n = 10000} \\
\hline
Model & Estimate & Emp. SD & Mean SE & Bias & Coverage 
      & Estimate & Emp. SD & Mean SE & Bias & Coverage \\
\hline

\multicolumn{11}{c}{$\beta_1$} \\
\hline
REF         & -0.246 & 0.089 & 0.083 & -0.006 & 0.95 
            & -0.243 & 0.043 & 0.046 & -0.003 & 0.95 \\
ApproxTIMM  & -0.187 & 0.089 & 0.082 &  0.054 & 0.87 
            & -0.176 & 0.048 & 0.045 &  0.065 & 0.74 \\
ApproxTIHMM & -0.212 & 0.104 & 0.096 &  0.028 & 0.92 
            & -0.200 & 0.057 & 0.052 &  0.040 & 0.85 \\
TIMM        & -0.192 & 0.088 & 0.082 &  0.048 & 0.88 
            & -0.182 & 0.047 & 0.045 &  0.058 & 0.77 \\
TIHMM       & -0.205 & 0.094 & 0.087 &  0.035 & 0.90 
            & -0.193 & 0.050 & 0.048 &  0.048 & 0.85 \\

\hline
\multicolumn{11}{c}{$\beta_2$} \\
\hline
REF         & -0.192 & 0.119 & 0.123 &  0.003 & 0.96 
            & -0.197 & 0.067 & 0.067 & -0.002 & 0.96 \\
ApproxTIMM  & -0.143 & 0.131 & 0.119 &  0.053 & 0.90 
            & -0.136 & 0.065 & 0.065 &  0.059 & 0.84 \\
ApproxTIHMM & -0.170 & 0.157 & 0.140 &  0.026 & 0.91 
            & -0.158 & 0.077 & 0.076 &  0.038 & 0.93 \\
TIMM        & -0.140 & 0.129 & 0.119 &  0.056 & 0.90 
            & -0.134 & 0.064 & 0.065 &  0.062 & 0.84 \\
TIHMM       & -0.151 & 0.138 & 0.126 &  0.044 & 0.91 
            & -0.143 & 0.069 & 0.069 &  0.053 & 0.84 \\

\hline
\multicolumn{11}{c}{$\beta_3$} \\
\hline
REF         & -0.007 & 0.077 & 0.090 & -0.007 & 0.97 
            & -0.002 & 0.043 & 0.049 & -0.002 & 0.97 \\
ApproxTIMM  & -0.005 & 0.085 & 0.089 & -0.005 & 0.97 
            & -0.004 & 0.054 & 0.049 & -0.004 & 0.92 \\
ApproxTIHMM & -0.007 & 0.101 & 0.104 & -0.007 & 0.96 
            & -0.005 & 0.061 & 0.057 & -0.005 & 0.92 \\
TIMM        & -0.005 & 0.086 & 0.089 & -0.005 & 0.97 
            & -0.004 & 0.053 & 0.049 & -0.004 & 0.92 \\
TIHMM       & -0.004 & 0.092 & 0.095 & -0.004 & 0.96 
            & -0.004 & 0.055 & 0.052 & -0.004 & 0.92 \\

\hline
 & \multicolumn{10}{c}{Irregular Follow-up} \\
\hline
 & \multicolumn{5}{c}{n = 3000} & \multicolumn{5}{c}{n = 10000} \\
\hline
Model & Estimate & Emp. SD & Mean SE & Bias & Coverage 
      & Estimate & Emp. SD & Mean SE & Bias & Coverage \\
\hline

\multicolumn{11}{c}{$\beta_1$} \\
\hline
REF         & -0.246 & 0.089 & 0.083 & -0.006 & 0.95 
            & -0.243 & 0.043 & 0.046 & -0.003 & 0.95 \\
ApproxTIMM  & -0.183 & 0.085 & 0.081 & 0.058 & 0.87 
            & -0.176 & 0.044 & 0.044 & 0.064 & 0.73 \\
ApproxTIHMM & -0.194 & 0.095 & 0.088 & 0.047 & 0.86 
            & -0.188 & 0.051 & 0.048 & 0.052 & 0.81 \\
TIMM        & -0.183 & 0.085 & 0.081 & 0.058 & 0.87 
            & -0.177 & 0.044 & 0.044 & 0.064 & 0.73 \\
TIHMM       & -0.189 & 0.089 & 0.084 & 0.051 & 0.86 
            & -0.183 & 0.047 & 0.046 & 0.058 & 0.76 \\

\hline
\multicolumn{11}{c}{$\beta_2$} \\
\hline
REF         & -0.192 & 0.119 & 0.123 & 0.003 & 0.96 
            & -0.197 & 0.067 & 0.067 & -0.002 & 0.96 \\
ApproxTIMM  & -0.140 & 0.120 & 0.117 & 0.056 & 0.89 
            & -0.131 & 0.067 & 0.064 & 0.065 & 0.80 \\
ApproxTIHMM & -0.149 & 0.135 & 0.128 & 0.047 & 0.90 
            & -0.140 & 0.073 & 0.069 & 0.056 & 0.85 \\
TIMM        & -0.140 & 0.120 & 0.117 & 0.055 & 0.89 
            & -0.131 & 0.067 & 0.064 & 0.065 & 0.80 \\
TIHMM       & -0.147 & 0.127 & 0.123 & 0.048 & 0.91 
            & -0.137 & 0.070 & 0.067 & 0.059 & 0.82 \\

\hline
\multicolumn{11}{c}{$\beta_3$} \\
\hline
REF         & -0.007 & 0.077 & 0.090 & -0.007 & 0.97 
            & -0.002 & 0.043 & 0.049 & -0.002 & 0.97 \\
ApproxTIMM  & -0.010 & 0.079 & 0.088 & -0.010 & 0.98 
            & -0.006 & 0.051 & 0.048 & -0.006 & 0.95 \\
ApproxTIHMM & -0.010 & 0.085 & 0.096 & -0.010 & 0.98 
            & -0.009 & 0.052 & 0.052 & -0.009 & 0.96 \\
TIMM        & -0.010 & 0.079 & 0.088 & -0.010 & 0.97 
            & -0.006 & 0.050 & 0.048 & -0.006 & 0.95 \\
TIHMM       & -0.010 & 0.080 & 0.092 & -0.010 & 0.98 
            & -0.006 & 0.052 & 0.050 & -0.006 & 0.95 \\
\end{tabular}
\end{threeparttable}
\end{centerbox}
\label{tab: transition 1}
\end{table}

\begin{table}[ht]
\caption{Simulation results for Transition 2 (From Mild MM to Death) }
\begin{centerbox}
\begin{threeparttable}
\setlength{\tabcolsep}{6pt}
\begin{tabular}{l c c c c c | c c c c c}

\hline
 & \multicolumn{10}{c}{Population Study (PS)} \\
\hline
 & \multicolumn{5}{c}{n = 3000} & \multicolumn{5}{c}{n = 10000} \\
\hline
Model & Estimate & Emp. SD & Mean SE & Bias & Coverage
      & Estimate & Emp. SD & Mean SE & Bias & Coverage \\
\hline

\multicolumn{11}{c}{$\beta_1$} \\
\hline
REF & -0.462 & 0.117 & 0.118 & -0.015 & 0.96 
                 & -0.453 & 0.060 & 0.064 & -0.006 & 0.97 \\
ApproxTIMM       & -0.475 & 0.138 & 0.135 & -0.028 & 0.94 
                 & -0.469 & 0.070 & 0.073 & -0.022 & 0.94 \\
ApproxTIHMM      & -0.481 & 0.146 & 0.142 & -0.034 & 0.93 
                 & -0.476 & 0.072 & 0.077 & -0.029 & 0.94 \\
TIMM             & -0.461 & 0.137 & 0.135 & -0.014 & 0.94 
                 & -0.455 & 0.068 & 0.073 & -0.008 & 0.95 \\
TIHMM            & -0.465 & 0.143 & 0.140 & -0.018 & 0.95 
                 & -0.458 & 0.070 & 0.076 & -0.011 & 0.95 \\

\hline
\multicolumn{11}{c}{$\beta_2$} \\
\hline
REF & 0.136 & 0.157 & 0.151 & -0.005 & 0.95 
                 & 0.135 & 0.081 & 0.082 & -0.006 & 0.95 \\
ApproxTIMM       & 0.127 & 0.184 & 0.173 & -0.014 & 0.94 
                 & 0.127 & 0.088 & 0.093 & -0.014 & 0.96 \\
ApproxTIHMM      & 0.125 & 0.191 & 0.181 & -0.016 & 0.94 
                 & 0.122 & 0.091 & 0.097 & -0.019 & 0.96 \\
TIMM             & 0.136 & 0.182 & 0.172 & -0.005 & 0.94 
                 & 0.135 & 0.086 & 0.093 & -0.006 & 0.96 \\
TIHMM            & 0.137 & 0.187 & 0.178 & -0.004 & 0.94 
                 & 0.134 & 0.090 & 0.096 & -0.007 & 0.96 \\

\hline
\multicolumn{11}{c}{$\beta_3$} \\
\hline
REF & 0.022 & 0.136 & 0.124 & 0.022 & 0.93 
                 & 0.001 & 0.062 & 0.067 & 0.001 & 0.96 \\
ApproxTIMM       & 0.012 & 0.163 & 0.142 & 0.012 & 0.93 
                 & 0.002 & 0.073 & 0.077 & 0.002 & 0.99 \\
ApproxTIHMM      & 0.011 & 0.172 & 0.149 & 0.011 & 0.92 
                 & 0.000 & 0.077 & 0.080 & 0.0002 & 0.98 \\
TIMM             & 0.011 & 0.160 & 0.141 & 0.011 & 0.92 
                 & 0.001 & 0.072 & 0.076 & 0.001 & 0.99 \\
TIHMM            & 0.010 & 0.167 & 0.147 & 0.010 & 0.90 
                 & 0.0005 & 0.075 & 0.079 & -0.0001 & 0.99 \\

\hline
 & \multicolumn{10}{c}{Irregular Follow-up (IV)} \\
\hline
 & \multicolumn{5}{c}{n = 3000} & \multicolumn{5}{c}{n = 10000} \\
\hline
Model & Estimate & Emp. SD & Mean SE & Bias & Coverage
      & Estimate & Emp. SD & Mean SE & Bias & Coverage \\
\hline

\multicolumn{11}{c}{$\beta_1$} \\
\hline
REF & -0.462 & 0.117 & 0.118 & -0.015 & 0.96 
                 & -0.453 & 0.060 & 0.064 & -0.006 & 0.97 \\
ApproxTIMM       & -0.469 & 0.140 & 0.138 & -0.022 & 0.96 
                 & -0.453 & 0.077 & 0.075 & -0.006 & 0.95 \\
ApproxTIHMM      & -0.470 & 0.145 & 0.141 & -0.023 & 0.96 
                 & -0.458 & 0.079 & 0.076 & -0.011 & 0.96 \\
TIMM             & -0.469 & 0.140 & 0.138 & -0.023 & 0.96 
                 & -0.453 & 0.076 & 0.075 & -0.006 & 0.95 \\
TIHMM            & -0.470 & 0.144 & 0.139 & -0.023 & 0.97 
                 & -0.454 & 0.077 & 0.075 & -0.008 & 0.96 \\

\hline
\multicolumn{11}{c}{$\beta_2$} \\
\hline
REF & 0.136 & 0.157 & 0.151 & -0.005 & 0.95 
                 & 0.135 & 0.081 & 0.082 & -0.006 & 0.95 \\
ApproxTIMM       & 0.137 & 0.178 & 0.176 & -0.004 & 0.95 
                 & 0.127 & 0.088 & 0.096 & -0.014 & 0.97 \\
ApproxTIHMM      & 0.137 & 0.184 & 0.180 & -0.003 & 0.95 
                 & 0.125 & 0.089 & 0.097 & -0.015 & 0.97 \\
TIMM             & 0.137 & 0.179 & 0.176 & -0.003 & 0.95 
                 & 0.126 & 0.088 & 0.095 & -0.014 & 0.97 \\
TIHMM            & 0.138 & 0.183 & 0.178 & -0.003 & 0.95 
                 & 0.125 & 0.089 & 0.096 & -0.015 & 0.97 \\

\hline
\multicolumn{11}{c}{$\beta_3$} \\
\hline
REF & 0.022 & 0.136 & 0.124 & 0.022 & 0.93 
                 & 0.001 & 0.062 & 0.067 & 0.001 & 0.96 \\
ApproxTIMM       & 0.016 & 0.165 & 0.145 & 0.016 & 0.91 
                 & 0.001 & 0.070 & 0.078 & 0.001 & 0.99 \\
ApproxTIHMM      & 0.015 & 0.169 & 0.148 & 0.015 & 0.91 
                 & 0.001 & 0.073 & 0.080 & 0.001 & 0.98 \\
TIMM             & 0.017 & 0.164 & 0.144 & 0.017 & 0.91 
                 & 0.001 & 0.069 & 0.078 & 0.001 & 0.99 \\
TIHMM            & 0.015 & 0.166 & 0.146 & 0.015 & 0.91 
                 & 0.001 & 0.070 & 0.079 & 0.001 & 0.98 \\

\hline
\end{tabular}
\end{threeparttable}
\end{centerbox}
\label{tab: transition 2}
\end{table}

\begin{table}[ht]
\caption{Simulation study results for Transition 3 (From Complex MM to Death)}
\begin{centerbox}
\begin{threeparttable}
\setlength{\tabcolsep}{6pt}
\begin{tabular}{l c c c c c | c c c c c}

\hline
 & \multicolumn{10}{c}{Population Study (PS)} \\
\hline
 & \multicolumn{5}{c}{n = 3000} & \multicolumn{5}{c}{n = 10000} \\
\hline
Model & Estimate & Emp. SD & Mean SE & Bias & Coverage
      & Estimate & Emp. SD & Mean SE & Bias & Coverage \\
\hline

\multicolumn{11}{c}{$\beta_1$} \\
\hline
REF & -0.305 & 0.142 & 0.139 &  0.005 & 0.93 
                 & -0.312 & 0.066 & 0.076 & -0.002 & 0.96 \\
ApproxTIMM       & -0.321 & 0.143 & 0.136 & -0.010 & 0.96 
                 & -0.323 & 0.066 & 0.074 & -0.013 & 0.97 \\
ApproxTIHMM      & -0.321 & 0.143 & 0.137 & -0.011 & 0.97 
                 & -0.325 & 0.066 & 0.075 & -0.015 & 0.96 \\
TIMM             & -0.312 & 0.140 & 0.136 & -0.002 & 0.96 
                 & -0.316 & 0.065 & 0.074 & -0.006 & 0.97 \\
TIHMM            & -0.312 & 0.139 & 0.136 & -0.002 & 0.97 
                 & -0.317 & 0.065 & 0.074 & -0.007 & 0.97 \\

\hline
\multicolumn{11}{c}{$\beta_2$} \\
\hline
REF & 0.153 & 0.181 & 0.190 & -0.034 & 0.93 
                 & 0.186 & 0.101 & 0.103 &  0.0001 & 0.95 \\
ApproxTIMM       & 0.134 & 0.187 & 0.186 & -0.052 & 0.97 
                 & 0.167 & 0.095 & 0.101 & -0.019 & 0.98 \\
ApproxTIHMM      & 0.132 & 0.188 & 0.187 & -0.054 & 0.95 
                 & 0.165 & 0.095 & 0.101 & -0.021 & 0.97 \\
TIMM             & 0.138 & 0.183 & 0.185 & -0.048 & 0.97 
                 & 0.171 & 0.096 & 0.100 & -0.015 & 0.98 \\
TIHMM            & 0.138 & 0.182 & 0.185 & -0.049 & 0.96 
                 & 0.171 & 0.096 & 0.100 & -0.015 & 0.98 \\

\hline
\multicolumn{11}{c}{$\beta_3$} \\
\hline
REF & -0.004 & 0.154 & 0.148 & -0.004 & 0.95 
                 &  0.006 & 0.083 & 0.080 &  0.006 & 0.93 \\
ApproxTIMM       &  0.013 & 0.154 & 0.146 &  0.013 & 0.95 
                 &  0.004 & 0.081 & 0.079 &  0.004 & 0.94 \\
ApproxTIHMM      &  0.015 & 0.156 & 0.146 &  0.015 & 0.97 
                 &  0.005 & 0.082 & 0.079 &  0.005 & 0.94 \\
TIMM             &  0.013 & 0.151 & 0.146 &  0.013 & 0.95 
                 &  0.005 & 0.080 & 0.079 &  0.005 & 0.93 \\
TIHMM            &  0.014 & 0.151 & 0.145 &  0.014 & 0.95 
                 &  0.005 & 0.081 & 0.079 &  0.005 & 0.93 \\

\hline
 & \multicolumn{10}{c}{Irregular Follow-up (IV)} \\
\hline
 & \multicolumn{5}{c}{n = 3000} & \multicolumn{5}{c}{n = 10000} \\
\hline
Model & Estimate & Emp. SD & Mean SE & Bias & Coverage
      & Estimate & Emp. SD & Mean SE & Bias & Coverage \\
\hline

\multicolumn{11}{c}{$\beta_1$} \\
\hline
REF & -0.305 & 0.142 & 0.139 &  0.005 & 0.93 
                 & -0.312 & 0.066 & 0.076 & -0.002 & 0.96 \\
ApproxTIMM       & -0.322 & 0.144 & 0.137 & -0.012 & 0.95 
                 & -0.328 & 0.075 & 0.075 & -0.018 & 0.93 \\
ApproxTIHMM      & -0.323 & 0.148 & 0.139 & -0.012 & 0.95 
                 & -0.328 & 0.074 & 0.076 & -0.018 & 0.94 \\
TIMM             & -0.322 & 0.145 & 0.137 & -0.012 & 0.94 
                 & -0.328 & 0.075 & 0.075 & -0.018 & 0.93 \\
TIHMM            & -0.325 & 0.148 & 0.139 & -0.015 & 0.95 
                 & -0.329 & 0.075 & 0.076 & -0.019 & 0.93 \\

\hline
\multicolumn{11}{c}{$\beta_2$} \\
\hline
REF & 0.153 & 0.181 & 0.190 & -0.034 & 0.93 
                 & 0.186 & 0.101 & 0.103 &  0.0001 & 0.95 \\
ApproxTIMM       & 0.143 & 0.175 & 0.186 & -0.043 & 0.94 
                 & 0.174 & 0.091 & 0.101 & -0.012 & 0.97 \\
ApproxTIHMM      & 0.141 & 0.179 & 0.190 & -0.045 & 0.93 
                 & 0.169 & 0.090 & 0.103 & -0.017 & 0.97 \\
TIMM             & 0.145 & 0.177 & 0.186 & -0.041 & 0.93 
                 & 0.175 & 0.092 & 0.101 & -0.011 & 0.97 \\
TIHMM            & 0.142 & 0.182 & 0.189 & -0.044 & 0.93 
                 & 0.175 & 0.093 & 0.102 & -0.011 & 0.97 \\

\hline
\multicolumn{11}{c}{$\beta_3$} \\
\hline
REF & -0.004 & 0.154 & 0.148 & -0.004 & 0.95 
                 &  0.006 & 0.083 & 0.080 &  0.006 & 0.93 \\
ApproxTIMM       &  0.007 & 0.151 & 0.147 &  0.007 & 0.93 
                 & -0.001 & 0.083 & 0.080 & -0.001 & 0.94 \\
ApproxTIHMM      &  0.009 & 0.154 & 0.149 &  0.009 & 0.93 
                 & -0.001 & 0.084 & 0.081 & -0.001 & 0.94 \\
TIMM             &  0.007 & 0.151 & 0.147 &  0.007 & 0.92 
                 & -0.001 & 0.083 & 0.080 & -0.001 & 0.94 \\
TIHMM            &  0.008 & 0.153 & 0.148 &  0.008 & 0.93 
                 &  -0.0001 & 0.083 & 0.081 &  -0.0001 & 0.94 \\

\hline
\end{tabular}
\end{threeparttable}
\end{centerbox}
\label{tab: transition 3}
\end{table}


%
\section*{Application to the SNAC-K Cohort}

SNAC-K is one of the 4 sites of the Swedish National study on Aging and Care (SNAC) \cite{SNAC}. SNAC is a longitudinal study on the elderly population, aimed at increasing knowledge about aging and health trends and providing a better basis for developing preventive measures and elder care \cite{SNAC}. 
At each follow-up, SNAC-K participants undergo a 5-hour-long comprehensive clinical and functional assessment carried out by trained physicians, nurses, and psychologists. Physicians collect information on diagnoses via physical examination, medical history, examination of medical charts, self-reported information, and/or proxy interviews. Clinical parameters, lab tests, drug information, and inpatient and outpatient care records are also used to identify specific conditions. Information on participants' deaths is derived from the cause of death register.
All diagnoses are coded in accordance with the International Classification of Diseases, 10th revision (ICD-10)\cite{ICD_10}. For the analyses conducted in this study, 60 chronic disease categories derived by a clinically
driven methodology reported in Calderón-Larra\~naga et al. \cite{CalderonLarranaga2017} were used.
The analyses were conducted on a subset of SNAC-K data, consisting of longitudinal records collected between 2001 and 2019 from participants enrolled at the start of the study (cohort 1). Follow-ups are carried out every 6 years for participants aged 60 to 78, and every 3 years for those aged 78 and older. At baseline, cohort 1 included 3,363 individuals. For this study, only participants with multimorbidity ($\ge$ 2 chronic conditions) at the first visit were included (N = 3,268). Participants with only one recorded visit (no follow-ups) or with missing data were excluded, resulting in a final sample of 2,716 individuals and a total of 9,085 observations. Over the 18-year follow-up period, 1,488 participants (55\%) died. The mean age at death was 85 years for men (sd=8.7) and 89 years for women (sd=8.0). At baseline, women represented 63\% of the total sample. Participants underwent between two and seven visits, with a median of 3 (IQR: 2-5) follow-ups per individual.

\begin{figure}[!ht]
    \centering
    \includegraphics[width=1\linewidth]{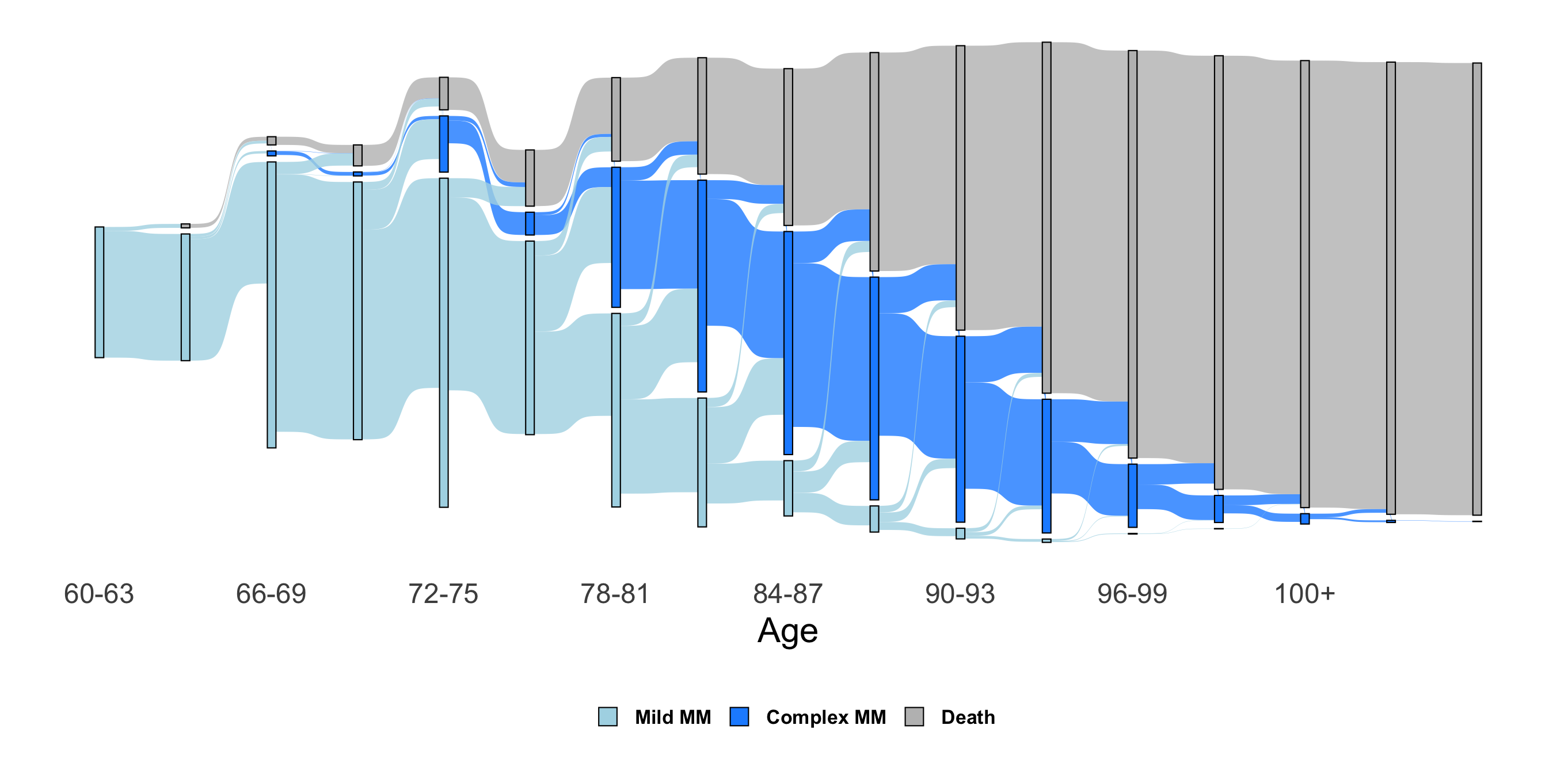}
    \caption{\textbf{ Alluvial diagram of observed transitions across multimorbidity states in the SNAC‑K cohort.} Each stream represents an individual’s assigned multimorbidity pattern over time (mild: light blue; complex: dark blue), with transitions to death shown in grey. Patterns are based on posterior membership probabilities from the latent class model and illustrate increasing movement toward complex multimorbidity with age.}
    \label{fig:Alluvial}
\end{figure}

Similarly to the simulation study, an illness-death multistate model is considered, containing two patterns of multimorbidity and death. The two patterns of multimorbidity are derived from Latent Class Analysis, which has been previously used to identify multimorbidity patterns in this population.  Conditional disease prevalences and their corresponding 95\% confidence intervals were used to interpret and label the two identified multimorbidity patterns (Table \ref{tab:lca}).

\begin{table}[ht]
\caption{Characterization of multimorbidity patterns by conditional disease prevalence (\%) and 95\% confidence intervals obtained from Latent Class model}
\centering
\begin{tabular}{l cc}
\hline
Disease & Mild MM & Complex MM \\
\hline
Anemia & 6.28 (5.08-7.73) & 29.98 (26.53-33.68) \\
Asthma & 7.45 (6.3-8.79) & 5.55 (3.99-7.68) \\
Atrial fibrillation & 3.64 (2.74-4.81) & 27.27 (23.92-30.9) \\
Autoimmune disease & 4.16 (3.31-5.22) & 7.22 (5.5-9.43) \\
Blindness visual loss & 0.86 (0.43-1.72) & 13.83 (11.4-16.68) \\
Bradycardias conduction disease & 0.42 (0.17-1.01) & 5.85 (4.35-7.83) \\
COPD, emphysema, chronic bronchitis & 3.73 (2.89-4.79) & 9.92 (7.84-12.49) \\
Cardiac valve disease & 1.34 (0.86-2.08) & 6.12 (4.56-8.16) \\
Cataract lens disease & 4.77 (3.79-5.98) & 9.5 (7.43-12.07) \\
Cerebrovascular disease & 3.71 (2.84-4.83) & 20.46 (17.62-23.62) \\
Chronic kidney disease & 29.84 (27.65-32.13) & 55.44 (51.26-59.54) \\
Colitis related disease & 8.01 (6.73-9.53) & 28.56 (25.17-32.22) \\
Deafness hearing loss & 6.5 (5.3-7.94) & 27.54 (24.22-31.12) \\
Dementia & 2.34 (1.53-3.54) & 29.59 (25.75-33.75) \\
Depression mood disease & 9.35 (7.99-10.92) & 12.83 (10.33-15.84) \\
Diabetes & 8 (6.76-9.44) & 14.28 (11.79-17.2) \\
Dorsopathies & 6.72 (5.62-8.02) & 8.58 (6.67-10.98) \\
Dyslipidemia & 59.9 (57.45-62.31) & 28.58 (24.99-32.45) \\
Esophageal, gastric and duodenal diseases & 4.7 (3.8-5.81) & 5.48 (4-7.48) \\
Glaucoma & 4.17 (3.27-5.31) & 11.24 (9.06-13.86) \\
Heart failure & 0.78 (0.34-1.81) & 36.84 (32.83-41.03) \\
Hypertension & 78.8 (76.73-80.73) & 61.15 (57.22-64.94) \\
Inflammatory arthropathies & 3.25 (2.47-4.26) & 7.49 (5.71-9.76) \\
Ischemic heart disease & 9.77 (8.34-11.4) & 34.54 (30.87-38.4) \\
Migraine and facial pain syndromes & 2.24 (1.64-3.05) & 3.37 (2.23-5.07) \\
Migraine and facial pain syndromes & 3.26 (2.47-4.3) & 4.18 (2.79-6.2) \\
Obesity & 15.88 (14.23-17.67) & 6.88 (5.07-9.27) \\
Osteoarthritis and degenerative joint diseases & 14.98 (13.37-16.74) & 12.58 (10.23-15.37) \\
Osteoporosis & 5.52 (4.47-6.79) & 12.43 (10.08-15.22) \\
Other MSK joint diseases & 5.68 (4.65-6.92) & 11.31 (9.15-13.9) \\
Other cardiovascular diseases & 0.91 (0.49-1.67) & 10.56 (8.39-13.22) \\
Other eye diseases & 4.19 (3.3-5.32) & 9.01 (7.03-11.47) \\
Other genitourinary diseases & 2.41 (1.78-3.25) & 3.88 (2.68-5.59) \\
Other neurological diseases & 1.98 (1.4-2.78) & 2.64 (1.65-4.2) \\
Other psychiatric and behavioral disorders & 1.16 (0.73-1.84) & 5.53 (4.07-7.46) \\
Prostate diseases & 3.99 (3.14-5.05) & 5.64 (4.11-7.69) \\
Sleep diseases & 2.73 (2.07-3.6) & 1.64 (0.89-2.99) \\
Solid neoplasms & 9.37 (8.06-10.87) & 11.71 (9.43-14.44) \\
Thyroid diseases & 11.45 (10.02-13.05) & 12.71 (10.38-15.46) \\
\hline
\end{tabular}
\label{tab:lca}
\end{table}

The "mild multimorbidity" pattern presents a higher prevalence of cardiovascular risk factors (hypertension, diabetes, dyslipidemia and obesity), asthma and sleep disorders; which have a lower risk of hospitalization and disability. On the other hand, the pattern denoted as"complex multimorbidity" is characterized by higher prevalence of cardiac diseases (heart failure, bradycardias conduction disease, atrial fibrillation, and anemia), dementia and sensory impairment diseases (deafness, hearing loss, blindness, and visual loss). The diseases characterizing the complex group are known to entail greater care needs and are associated with poorer physical and cognitive function.
Based on this classification, individuals' transitions across multimorbidity states are modeled using the hidden multistate model implemented in the \texttt{nhm} R package, which demonstrated the best performance in the simulation study.  We include in the analysis a set of potential lifestyle and socio-economic risk factors that may increase the risk of transitioning towards the complex multimorbidity state, as well as death. Specifically, baseline covariates included in the multistate model are sex, level of education, physical activity, living alone, alcohol consumption, smoking,  and manual occupation.

Out of the 2,716 participants, at baseline 1,864 (69\%)  individuals  are assigned to the mild multimorbidity state and 852 (31\%) to the complex multimorbidity state. The alluvial plot shown in Figure \ref{fig:Alluvial} describes how individuals in the SNAC-K cohort transition among the two multimorbidity states and death over the older age course. As participants grow older, transitions toward complex multimorbidity become increasingly frequent. Moreover, individuals in the complex multimorbidity state exhibit higher mortality rates than those in the mild state of the same age.

\begin{figure}
    \centering
    
    \begin{subfigure}[h]{0.53\textwidth}
        \centering
        \includegraphics[width=\textwidth]{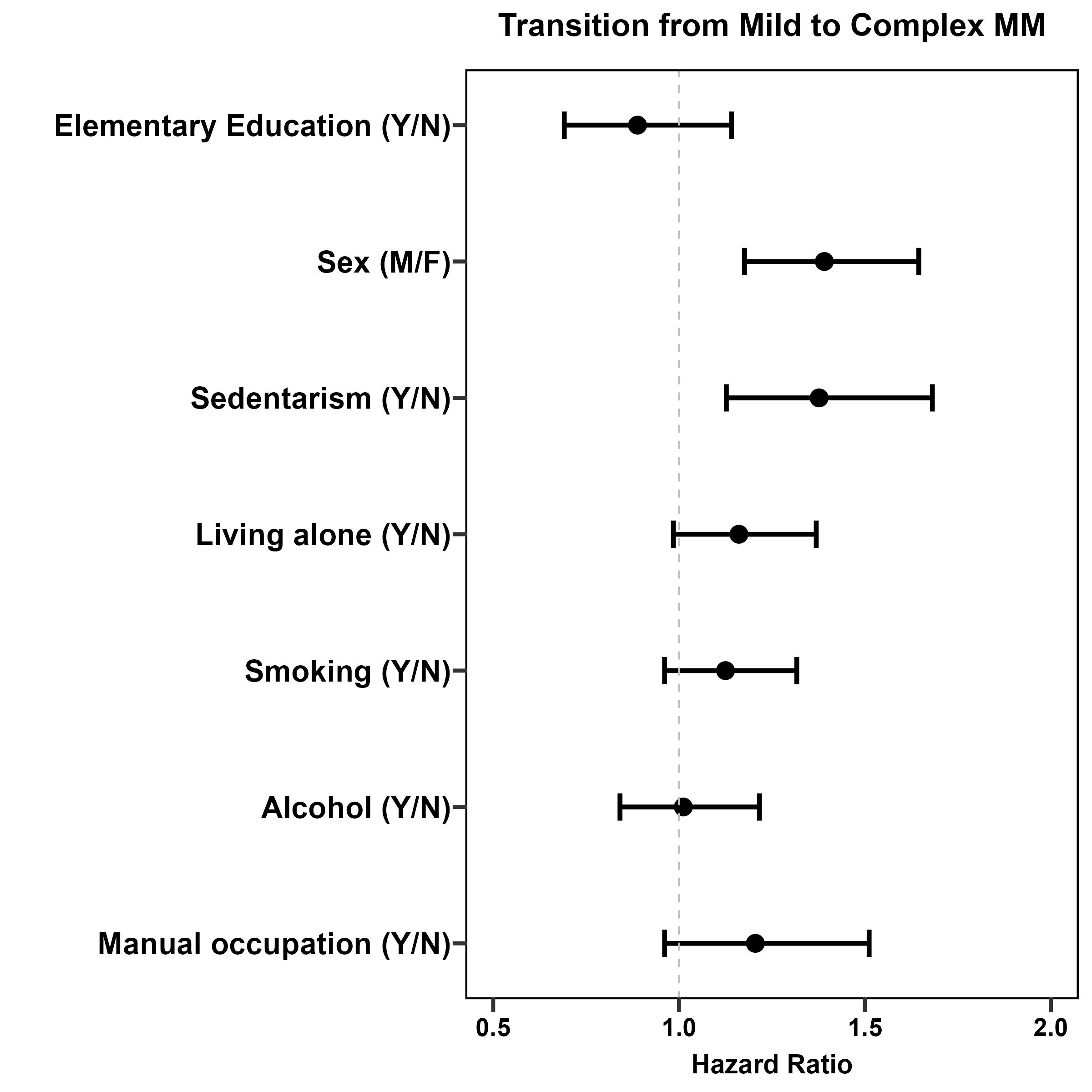}
        \caption{Hazard ratios with 95\% confidence intervals for transitioning from mild to complex multimorbidity, estimated using the hidden multistate model. }
        \label{fig:HR}
    \end{subfigure}%
    \hfill
    \begin{subfigure}[h]{0.43\textwidth}
        \centering
        \includegraphics[width=\textwidth]{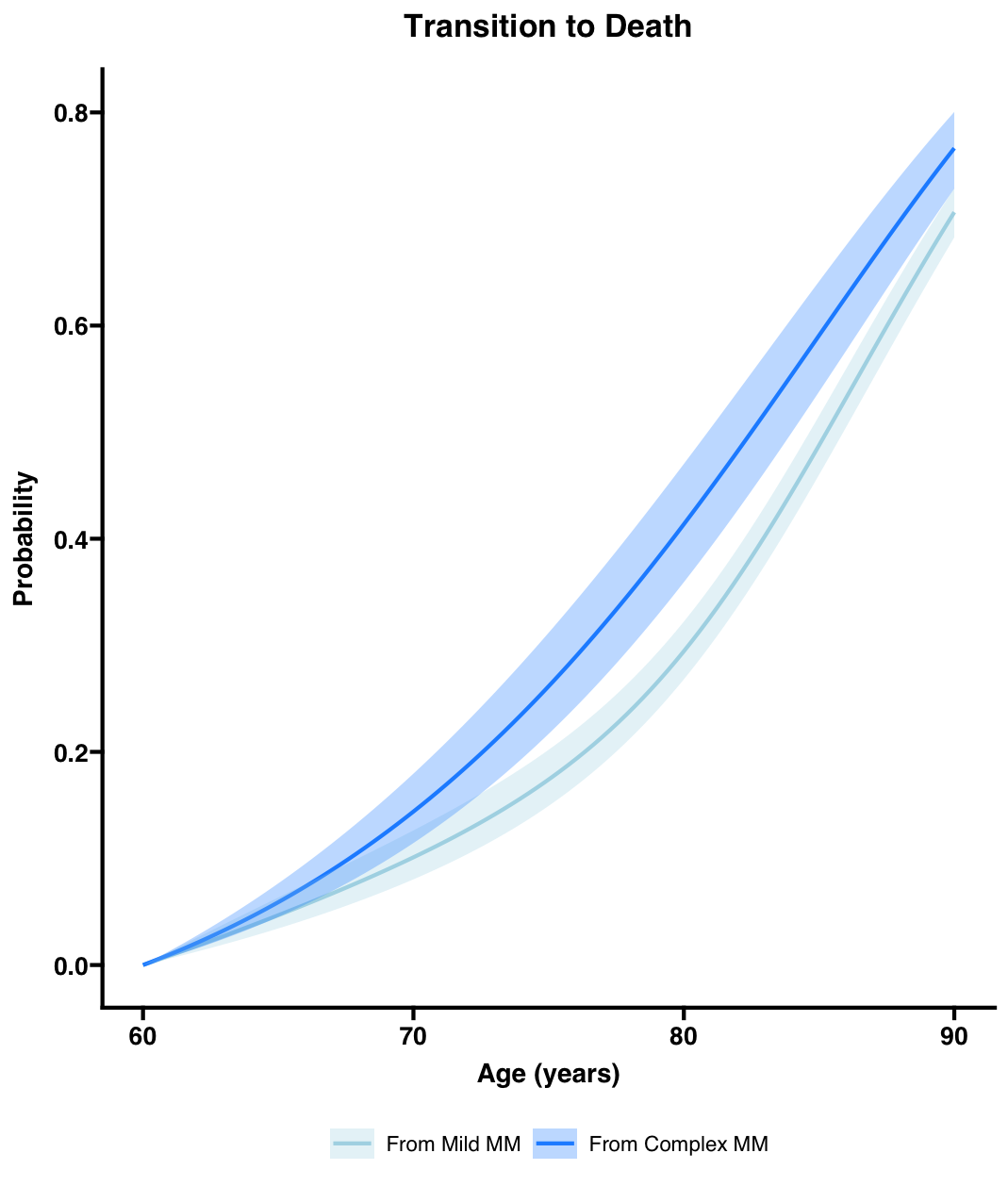}
        \caption{Predicted probability of death from age 60 by multimorbidity state, showing higher mortality risk for individuals in the complex state; all covariates were held at cohort mean values. }
        \label{fig:death_transitions_by_state}
    \end{subfigure}

    \begin{subfigure}[h]{\textwidth}
        \centering
        \includegraphics[width=\textwidth]{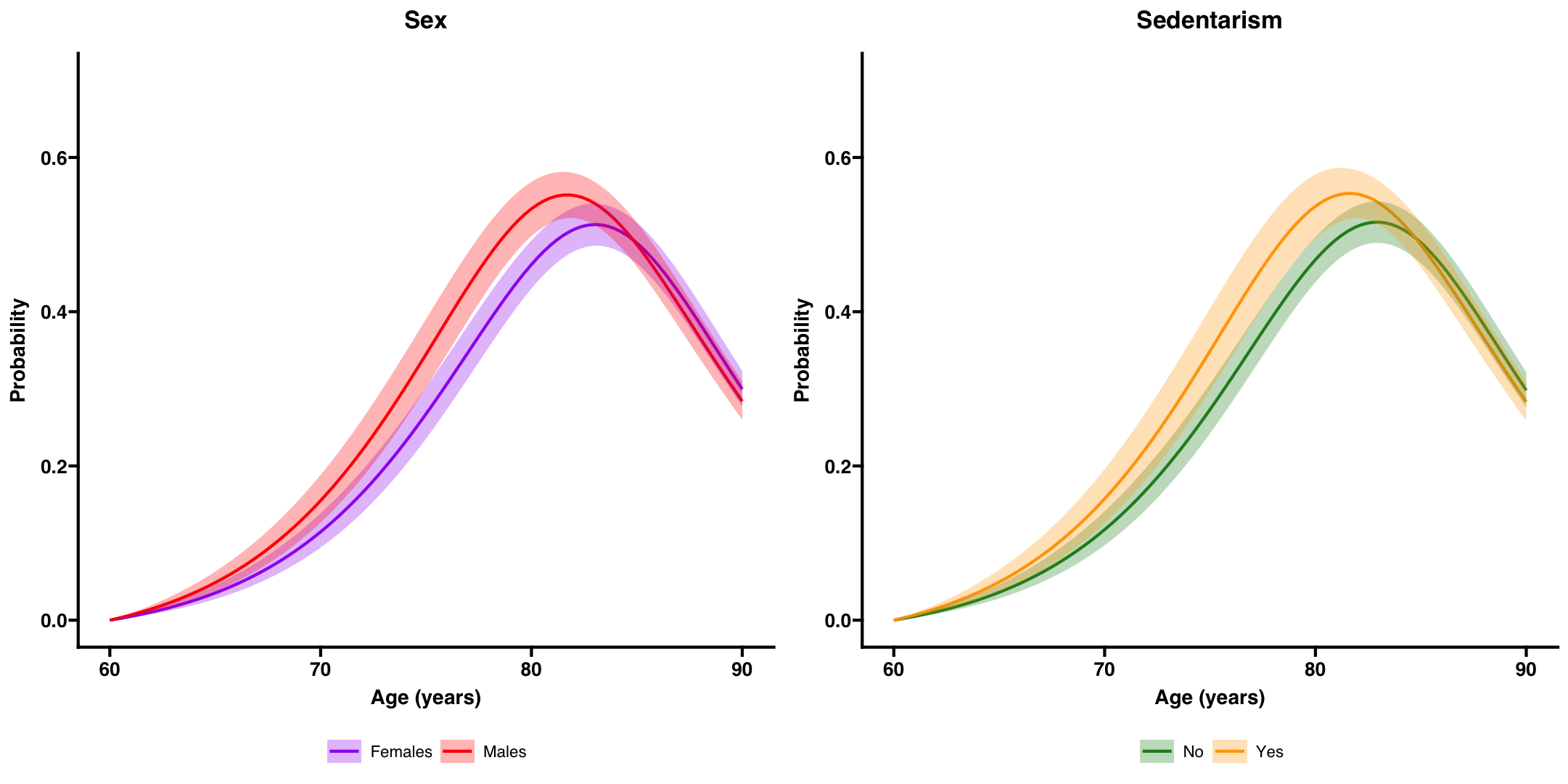}
        \caption{Predicted probability of progressing to complex multimorbidity from age 60, comparing males vs. females (\textbf{left panel}) and sedentary vs. non‑sedentary behavior (\textbf{right panel}), with all other covariates held at cohort mean values. }
        \label{fig: 4c}
    \end{subfigure}

\end{figure}

Hazard ratios for the transition from mild to complex multimorbidity, as estimated by the hidden multistate model, are presented in Figure \ref{fig:HR}. Sedentary behavior and being male are associated with a higher hazard of transitioning to complex multimorbidity. Figure \ref{fig: 4c} illustrates how these findings translate into the predicted probability of transitioning to complex multimorbidity, comparing males versus females (right panel) and sedentary versus non-sedentary behavior (left panel) from age 60, while holding other exposures at their cohort mean values. In both panels, the higher-risk groups exhibit a greater probability of progressing to the complex multimorbidity state up to around age 85, after which the increasing risk of death reduces the observed differences. Finally, the predicted probability of dying from the two different multimorbidity states is compared in Figure \ref{fig:death_transitions_by_state}, confirming the higher mortality associated with complex multimorbidity.

\section*{Discussion}

In this work, we investigated the use of continuous-time hidden markov models to study transitions between multimorbidity patterns. Through a comprehensive simulation study, we demonstrated the potential of these models across various scenarios. Additionally, by applying the model to real-world data, we illustrated its practical utility in multimorbidity research.
The current literature on transitions across multimorbidity patterns primarily relies on discrete time approaches. For instance, Roso Llorach, Vetrano et al. \cite{AlbertVetrano} employed discrete-time Hidden Markov models to investigate transitions between multimorbidity clusters in the SNAC K dataset. A key distinction from our approach is that their analysis stratified participants by baseline age group, fitting separate discrete time models for each group. Although this method captures age-related heterogeneity, it prevents the use of age as a continuous timescale and complicates interpretation, as different cluster sets may arise for the same age group at different follow-up times. Moreover, their approach does not allow the estimation of covariate effects, limiting the identification of risk profiles associated with different transition patterns.
Zacarias Pons et al. \cite{zacarias2021multimorbidity} applied Latent Transition Analysis (LTA) \cite{collins2010latent} to self-reported multimorbidity data from the SHARE study. LTA enables modeling transitions between latent classes and incorporating covariates, but transition probabilities must be re-estimated for each interval, making the method less suitable for datasets with many time points or irregular follow-ups. Importantly, in both approaches, right censoring and survival processes cannot be modeled according to survival analysis standards.

In contrast, the continuous-time framework proposed in this study addresses these limitations and is better suited to complex longitudinal study designs. Hidden multistate models naturally handle irregular follow-up intervals, time to death, and censoring mechanisms, and allow modeling of time-varying transition hazards. They also enable the estimation of covariate effects on transitions — a key feature for identifying risk factors for progression within multimorbidity trajectories. Our simulation served a dual purpose: it enabled both the comparison of different model specifications and the evaluation of the reliability of the proposed framework. The results underscored the importance of incorporating the latent nature of the states and a fully time-inhomogeneous baseline hazard. In some scenarios, confidence-interval coverage fell below nominal levels. This likely reflects limitations of Hessian-based variance estimation, suggesting that bootstrap methods may provide more robust uncertainty quantification. Beyond the simulation setting, the application to the SNAC K cohort confirmed that the modeling framework can deliver clinically meaningful insights in real-world epidemiological settings. By integrating latent multimorbidity patterns with continuous-time transition modeling, the analysis identified relevant sociodemographic and behavioral factors associated with progression to complex multimorbidity, and highlighted the marked differences in mortality risk between patterns. Although we focused on two multimorbidity states—since the aim of this paper is to demonstrate the applicability of the method rather than to provide novel substantive evidence in the context of multimorbidity—the modeling framework is inherently flexible and capable of accommodating more than two states. This scalability will be crucial in future epidemiological studies applying this method.

Model identifiability is also a key consideration in hidden multistate models \cite{titman2011flexible}, as formal identifiability results are generally not available for models of this complexity. In this study, we mitigated these challenges through a parsimonious and structured modeling strategy. In particular, transition intensities were specified using a parametric Gompertz formulation, ensuring a flexible yet low-dimensional representation of age-dependent dynamics, and the separation between the latent class model (defining multimorbidity patterns) and the hidden multistate model (describing their evolution) avoided overparameterization of the emission process, which is estimated from the latent class model and then treated as fixed in the likelihood of the hidden multistate model. 
In addition, neither in the simulation study nor in the application did we encounter problems with the observed information matrix being singular, which is considered the definition for a model to be empirically identifiable \cite{latentbook}.  Nonetheless, identifiability should remain a central consideration in future applications. In particular, careful attention should be paid to: (i) the definition and interpretability of multimorbidity patterns, ensuring that latent states are well separated (e.g., high diagonal dominance in the emission matrix); (ii) the trade-off between the number of latent states, model flexibility, and available sample size; and (iii) the empirical assessment of identifiability, for instance through simulation or sensitivity analyses. When numerical issues arise—such as singularity of the observed information matrix—this should be taken as an indication to revise the model specification, for example by simplifying the state structure or introducing additional constraints.

Two more assumptions of the methodology are also important to mention in relation to model validity. First, a well-defined clustering or latent model must be available to ensure that the states also represent clinically meaningful and homogeneous multimorbidity patterns. Second, the set of states must encompass all relevant patterns that may appear over time. Therefore, when using this methodology, it is essential to identify clinically meaningful multimorbidity patterns in a sample that is representative of all age groups of interest. This aligns with current recommendations in the multimorbidity trajectory literature \cite{Nagel2024,CalderonLarranaga2025}.
\section*{Conclusion}

In this work, we presented a modeling strategy that integrates latent multimorbidity patterns with continuous‑time hidden multistate models to study the dynamic evolution of multimorbidity. The simulation study demonstrated that accounting for both the latent nature of the states and time‑inhomogeneous transition hazards is essential for reliable inference, and the application to the SNAC‑K cohort illustrated how this framework can yield clinically meaningful insights in real‑world settings.
Although motivated by multimorbidity, the framework is broadly applicable to longitudinal processes characterized by latent constructs, irregular observation schedules, and heterogeneous progression dynamics. Many aging‑related phenomena—such as cognitive decline, disability progression, frailty, or evolving care needs—share these features. As such, the proposed approach provides a general and versatile tool for modeling complex health trajectories, with the potential to support risk stratification and inform preventive strategies in aging research.

\section*{Data availability}
The simulated data were generated using the code available at \url{https://github.com/ARCbiostat/SimAgingData.git}, and the code used for the analysis is available at \url{https://github.com/ARCbiostat/CTHMM_multimorbidity}.
Access to original SNAC-K data is available to the research community upon approval by the SNAC-K data management and maintenance committee. Applications for accessing these data can be submitted through the website (https://www.snac-k.se/).

\section*{Funding}
F.I. acknowledges the support by MUR, grant Dipartimento di Eccellenza 2023-2027. C.G. acknowledges the support by the Loo och Hans Ostermans Stiftelse för medicinsk forskning.  D.L.V. 
acknowledges the support by the Swedish Research Council (project number 
2021-03324), the Karolinska Institutet Strategic Research Area in 
Epidemiology and Biostatistics in 2021 and 2023, and the Karolinska 
Institutet Strategic Research Area in Neuroscience in 2025. 

\bibliography{sample}

\section*{Acknowledgements}

We thank the SNAC-K participants and the SNAC-K
Group for their collaboration in data collection and management. 

\section*{Author contributions statement}

\textbf{Conceptualization}: V.M., C.G.\\
\textbf{Methodology}: V.M., C.G., F.I., D.L.V.\\
\textbf{Software / Analytical Pipeline Development}: V.M., C.G.\\
\textbf{Formal Analyses}: V.M., C.G.\\
\textbf{Writing – Original Draft}: V.M., C.G.\\
\textbf{Writing – Review and Editing}: F.I., D.L.V., A.C.L., V.M., C.G.\\
\textbf{Interpretation of Findings}: All authors.\\
\textbf{Approval of Final Manuscript}: All authors.





\end{document}